%Paper: 9202028
%From: ahn@strange.tn.cornell.edu (Changrim Ahn)
%Date: Fri, 7 Feb 92 17:01:49 EST

%%%%%%%%%%%%%%%%%%%%%%%%%%%%%%%%%%%%%%%%%%%%%%%%%%%%%%%%%%%%%%%%%%%%%%%
%%%% Macropackage=PHYZZX %%%%%%%%%%%%%%%%%%%%%%%%%%%%%%%%%%%%%%%%%%%%%%
%%%%%%%%%%%%%%%%%%%%%%%%%%%%%%%%%%%%%%%%%%%%%%%%%%%%%%%%%%%%%%%%%%%%%%%
\titlepage
\hfill{CLNS 92/1135\par}
\title{\bf RG Flows of Non-Unitary Minimal CFTs}
\author{Changrim Ahn
\foot{E-mail address:ahn@cornella.bitnet}  }
\address{\rm F. R. Newman Lab. of Nuclear Studies\break
          Cornell University\break
          Ithaca, NY 14853}
\abstract{
In this paper we study the renormalization group flow of the
$(p,q)$ minimal (non-unitary) CFT perturbed by the $\Phi_{1,3}$
operator with a positive coupling.
In the perturbative region $q>>(q-p)$,
we find a new IR fixed point which corresponds to the $(2p-q,p)$ minimal
CFT. The perturbing field near the new IR fixed point is identified with
the irrelevent $\Phi_{3,1}$ operator.
We extend this result to show that
the non-diagonal ($(A,D)$-type) modular invariant
partition function of the $(p,q)$ minimal CFT flows into the $(A,D)$-type
partition function of the $(2p-q,p)$ minimal CFT and the diagonal
partition function into the diagonal.}

\endpage
%%%%%%%%%%%%%%%%%%%%%%%%%%%%%%%%%%%%%%%%%%%%%%%%%%%%%%%%%%%%%%%%%%%%%%%
\def\bz{{\overline z}}
\def\btau{{\overline \tau}}

\def\ch{{\raise3pt\hbox{$\font\bigtenrm=cmr10 scaled\magstep1\chi$}}}
%%%%%%%%%%%%%%%%%%%%%%%%%%%%%%%%%%%%%%%%%%%%%%%%%%%%%%%%%%%%%%%%%%%%%
\REF\BPZ{A.A. Belavin, A.M. Polyakov, and A.B. Zamolodchikov,
Nucl. Phys. {\bf B241} (1984) 333.}
\REF\zamoli{A.B. Zamolodchikov, Int. J. Mod. Phys. {\bf A4} (1989)
4235.}
\REF\zamolii{A.B. Zamolodchikov, JETP Lett. {\bf 43} (1986) 702;
Sov. J. Nucl. Phys. {\bf 46} (1987) 1090.}
\REF\cardy{A.W.W. Ludwig and J.L. Cardy, Nucl. Phys. {\bf B285}
(1987) 687.}
\REF\FQS{D. Friedan, Z. Qiu, and S. Shenker, Phys. Rev. Lett. {\bf 52}
(1984) 1575.}
\REF\rsg{A. LeClair, Phys. Lett. {\bf 230B} (1989) 103;\hfil\break
D. Bernard and A. LeClair, Phys. Lett. {\bf 247B} (1990) 309.}
\REF\RaS{F. Smirnov, Int. J. Mod. Phys. {\bf A4} (1989) 4213;\hfil\break
N.Yu Reshetikhin and F. Smirnov, Comm. Math. Phys. {\bf 131} (1990) 17.}
\REF\CIZ{A. Cappelli, C. Itzykson, and J.-B. Zuber, Nucl. Phys.
{\bf B280[FS18]} (1987) 445.}
\REF\GaS{D. Ghoshal and A. Sen, Phys. Lett. {\bf B265} (1991) 295.}
\REF\Rava{F. Ravanini, Saclay Preprint SPhT/91-147 (1991).}
\REF\Elit{S. Elitzur, A. Giveon, and E. Rabinovici, Nucl. Phys. {\bf
B316} (1989) 679.}
\REF\DaF{Vl.S. Dotsenko and V.A. Fateev, Phys. Lett. {\bf B154}
(1985) 291.}
\REF\DSZ{P. di Francesco, H. Saleur, and J.-B. Zuber, J. Stat. Phys.
{\bf 49} (1987) 57.}
\REF\TBA{Al.B. Zamolodchikov, Nucl. Phys. {\bf B342} (1990) 695.}
\REF\zami{Al.B. Zamolodchikov, Nucl. Phys. {\bf B358} (1991) 619.}
\REF\AaN{C. Ahn and S. Nam, Phys. Lett. {\bf B271} (1991) 329.}
\REF\ahn{C. Ahn, Cornell Preprint 91/1117.}
\REF\AaNi{C. Ahn and S. Nam, in preparation.}
\REF\Cardy{J.L. Cardy, Phys. Rev. Lett. {\bf 54} (1985) 1354.}
\REF\Coset{C. Crnkovic, G.M. Sotkov, and M. Stanishkov,
Phys. Lett. {\bf 226} (1989) 297.}
\REF\ahni{C. Ahn, Nucl. Phys. {\bf B354} (1991) 57.}
\REF\lassig{M. Lassig, PRINT-92-0009 (JULICH) December 1991.}
%%%%%%%%%%%%%%%%%%%%%%%%%%%%%%%%%%%%%%%%%%%%%%%%%%%%%%%%%%%%%%%%%%%

{\bf 1.} Conformal field theories (CFTs)\refmark\BPZ\
perturbed by relevent
operators have provided a theoretical framework for constructing
integrable scale non-invariant 2D quantum field theories (QFTs).
There are two interesting classes:
One is a class of massive integrable QFTs
whose scattering matrices are exactly solvable due to an
infinite number of conserved currents.
\refmark\zamoli\
The other is a class of scale non-invariant QFTs with no massive particles
which have new RG fixed
points at which
the scale invariance and conformal symmetries are
restored.\refmark{\zamolii,\cardy}\

The minimal CFTs\refmark\BPZ\ are
characterized by two coprime integers $p,q$ $(q>p)$. We will denote the
$(p,q)$ minimal CFT by ${\cal M}_{(p,q)}$.
The principal series of $(p,p+1)$ correspond to the unitary CFTs in the
sense that the states created by the Virasoro generators have positive
definite norm.\refmark\FQS\
Except for the string theory, the unitarity of these
Virasoro modules seems not the first principle to be satisfied.
Interesting applications of the non-unitary CFTs have been made in the
integrable lattice models, matrix models and others.

We start with the following perturbed CFT:
$${\cal S}_{(p,q)}(g)={\cal M}_{(p,q)}+g\int d^2
z \Phi_{1,3}(z,\bz),\eqn\eq$$
where the dimension of the least relevent operator
$\Phi_{1,3}$ of ${\cal M}_{(p,q)}$ is
$$\Delta(\Phi_{1,3})=1-{2(q-p)\over{q}}.\eqn\dimen$$
The theories can be represented in one-dimensional parameter space spanned
by $g$ because $\Phi_{1,3}$ satisfies a closed
operator product expansion,
$$[\Phi_{1,3}]\times[\Phi_{1,3}]=[\Phi_{1,3}].\eqn\eq$$

If $g<0$, the theory is a massive integrable QFT, the
`restricted sine-Gordon theory' both for the unitary CFTs\refmark\rsg\
and the non-unitary CFTs.\refmark\RaS\
If $g>0$, the perturbed theory remains as a massless theory while
the scale invariance is broken.
For the unitary CFT ${\cal M}_{(p,p+1)}$,
A.B. Zamolodchikov\refmark\zamolii\ and A. Ludwig and J. Cardy\refmark\cardy\
found a new IR fixed point corresponding to another unitary theory ${\cal
M}_{(p-1,p)}$ for $\Delta(\Phi_{1,3})=1-\epsilon$ with $\epsilon=
2/(p+1)<<1$.
The perturbing field $\Phi_{1,3}$ becomes
irrelevent field $\Phi_{3,1}$ near the new fixed point.

In this paper, we study the RG flows of
the non-unitary CFTs ${\cal M}_{(p,q)}$ to show that
${\cal M}_{(p,q)}$ perturbed by $\Phi_{1,3}$ flows into
${\cal M}_{(2p-q,p)}$ perturbed by $\Phi_{3,1}$
for $q>>(q-p)$.
We also investigate the RG flows of the non-diagonal modular invariant
partition functions (MIPFs) of ${\cal M}_{(p,q)}$\refmark\CIZ\
to find that the $(A,A)$ and $(A,D)$ MIPFs flow into the
$(A,A)$ and $(A,D)$ MIPFs, respectively.
This analysis has been already made for the unitary CFTs in
[\GaS,\Rava].

{\bf 2.} Under the scale transformation $x_{\mu}\to (1+1/2 dt)x_{\mu}$,
the parameter $g$ changes according to the
equation\refmark\zamolii\
$$\beta(g)={dg\over{dt}}=\epsilon g-{1\over{2}}C
g^2 +{\cal O}(g^3),\quad \epsilon= {2(q-p)\over{q}}\eqn\rgeq$$
where the coefficient of a (diagonal) three-point function
$C=C_{(1,3)(1,3)(1,3)}$ is\refmark\DaF\
$$C_{(1,3)(r,s)(r,s)}=-{\Gamma(\epsilon)\over{\Gamma(1-\epsilon)}}
\left[{\gamma^3(1-\epsilon/2)\over{\gamma(2-3\epsilon/2)}}\right]^{1/2}
{\gamma(r+1-s-(1-s)\epsilon)\over{\gamma(r-s+(1+s)\epsilon)}},\eqn\coeff$$
with $\gamma(x)=\Gamma(x)/\Gamma(1-x)$.
For $\epsilon<<1$, $C\cong 4/\sqrt{3}+{\cal O}(\epsilon)$.

{}From Eq.\rgeq, one can find a new fixed point at $g=g_{*}$
$$\beta(g_{*})=0\longrightarrow g_{*}\cong
{\sqrt{3}\over{2}}{2(q-p)\over{q}}.\eqn\fixed$$
One can identify this as the IR fixed point from the fact that $g_{*}
=g(t_{*})$ with $t_{*}=\infty$ from Eq.\rgeq.
Following [\zamolii], we introduce the `$c$-function'
which gives the central charges at the fixed points and
satisfies the following RG equation:
$${dc\over{dt}}=\beta(g){\partial\over{\partial g}}c(g)=-12\beta^2(g).
\eqn\crgeq$$
Eq.\crgeq\ shows that the $c$-function is monotonically decreasing as
$t$ increases in the $\Phi_{1,3}$ direction such that
the inequality $c_{\rm\scriptscriptstyle UV}>c_{\rm\scriptscriptstyle
IR}$ holds.
\foot{
This does not contradict to the fact that `$c$-theorem'\refmark\zamolii\
is not valid for the non-unitary CFTs. Although the $c$-function is
not monotonically decreasing for {\it any} relevent perturbation, there
can still exist certain perturbations which makes $c$ decrease.}
{}From Eqs.\rgeq\ and \crgeq, the central charge of the IR CFT is given
by
$$\eqalign{c_{\rm\scriptscriptstyle IR}=
c(g_{*})&=c_{\rm\scriptscriptstyle UV}
-6\epsilon g_{*}^2 + 2C g_{*}^3 +{\cal
O}(g_{*}^4),\quad c_{\rm\scriptscriptstyle UV}=
1-{6(q-p)^2\over{pq}}\cr
&=1-{6(q-p)^2\over{p(2p-q)}}=
c[{\cal M}_{(2p-q,p)}].\cr}\eqn\IRcharge$$

The conformal dimension of the perturbing field near the new fixed point
is determined by
$$\eqalign{\Delta(\Phi_{1,3})&=1-{\partial\beta\over{\partial g}}
\biggm\vert_{g_{*}}
=1+\epsilon+\epsilon^2+{\cal O}(\epsilon^3)\cr
&\cong{q\over{2p-q}}=\Delta[\Phi_{3,1}]\quad{\rm for}\quad {\cal
M}_{(2p-q,p)}.\cr}\eqn\dimeni$$
This completes the RG flow of ${\cal M}_{(p,q)}$ perturbed by
$\Phi_{1,3}$ into ${\cal M}_{(2p-q,p)}$ perturbed by $\Phi_{3,1}$
for $q>>(q-p)$.
This means that the $(p,q)$ minimal CFTs can be grouped
into an infinite number of series for each value of $q-p$ such that the
CFTs in each series are connected by the RG flows under the $\Phi_{1,3}$
perturbations:
$$\cdots\to(p+n,p+2n)\to(p,p+n)\to(p-n,p)\to(p-2n,p-n)\to\cdots.$$
for $p>>n$ and $n=1,2,...$. The $n=1$ case gives the RG flows of the
principal series (the unitary CFTs). No unitary CFTs flow into the
non-unitary ones and vice versa.
This result has been also considered in ref.[\Elit].

{\bf 3}. We now consider how each MIPF of the minimal CFTs will flow
under the perturbation.
There are three types of the MIPFs for the $(p,q)$ minimal CFT, denoted
by $Z_{(p,q)}(A,A)$, $Z_{(p,q)}(A,D)$, and $Z_{(p,q)}(A,E)$.\refmark\CIZ\
The diagonal MIPFs $Z_{(p,q)}(A,A)$ are possible for all $(p,q)$ and
contains only spinless fields.
The non-diagonal MIPFs $Z_{(p,q)}(A,D)$ are possible for $p$ or $q$ even
and include some primary fields with spins.
The exceptional MIPFs $Z_{(p,q)}(A,E)$ are allowed
only for a few discrete values of $p$ and $q$.
We will consider the RG flows of the $Z_{(p,q)}(A,A)$ and
$Z_{(p,q)}(A,D)$ MIPFs here.

The partition functions of the perturbed $(p,q)$ minimal CFTs on the RG
trajectory can be written in the following schematic form:
$$Z(g)=\int[{\cal D}\varphi] e^{-{\cal S}_{(p,q)}(g)[\varphi]}
=\int[{\cal D}\varphi]\exp-\left[{\cal M}_{p/q}+g\int d^2
z \Phi_{1,3}(z,\bz)\right],\eqn\part$$
where $\varphi$ denotes the field degree of freedom which has no direct
consequence in the following context.
For a small $g_{*}<<1$, Eq.\part\ gives with $Z(0)=Z_{(p,q)}$ and
$Z(g_{*})=Z_{(2p-q,p)}$,
$$\delta Z=Z_{(2p-q,p)}(A,Y)-Z_{(p,q)}(A,X)
\cong -g_{*}{\tau_2\over{\pi}}\big\langle\Phi_{1,3}
\big\rangle_{\rm torus},\eqn\differ$$
where $\tau_2$ is the imaginary part of the modular parameter $\tau$.
For a given MIPF of ${\cal M}_{(p,q)}$ ($X=A$ or $D$), we need to find
the corresponding $Y$ for the MIPFs of ${\cal M}_{(2p-q,p)}$.
We will show that $Y=A$ if $X=A$ and $Y=D$ if $X=D$ using Eq.\differ.

For the purpose, it is convenient to express the MIPFs
of the minimal CFTs in terms of the Coulomb gas form:\refmark\DSZ\
$$\eqalign{&Z_{(p,q)}(A,A)={1\over{2}}\left[
Z_c\left(pq\right)-Z_c\left({p\over{q}}\right)\right],\cr
&Z_{(p,q)}(A,D)={1\over{2}}\left[
Z_c\left({4p\over{q}}\right)-Z_c\left({p\over{q}}\right)
-Z_c\left({4\over{pq}}\right)
+Z_c\left({1\over{pq}}\right)\right],\cr
&\qquad Z_c(N)={1\over{|\eta(\tau)|^2}}\sum_{e,m\in Z}
e^{\pi i\tau(e/\sqrt{N} +m\sqrt{N})^2/2}
e^{-\pi i\btau(e/\sqrt{N} -m\sqrt{N})^2/2}.\cr}\eqn\parti$$

We first compute $\delta Z$ for the diagonal and non-diagonal MIPFs
separately assuming that $(A,A)$ flows into $(A,A)$ and
$(A,D)$ to $(A,D)$:
$$\eqalign{\delta_A Z&=Z_{(2p-q,p)}(A,A)-Z_{(p,q)}(A,A)\cr
&\cong 2\pi\tau_2 {(q-p)\over{q}}
{1\over{|\eta(\tau)|^2}}F(pq)+{\cal O}(\epsilon^2),\cr
\delta_D Z&=Z_{(2p-q,p)}(A,D)-Z_{(p,q)}(A,D)\cr
&\cong 2\pi\tau_2 {(q-p)\over{q}}
{1\over{|\eta(\tau)|^2}}\left[F(pq)-F(pq/4)\right]
+{\cal O}(\epsilon^2),\cr}\eqn\eq$$
using $Z_c(N)=Z_c(1/N)$ and
$$\eqalign{&\delta Z_c(N)=Z_c(N+\delta N)-Z_c(N)\cong -2\pi\tau_2
{\delta N\over{N}}{1\over{|\eta(\tau)|^2}}F(N)\cr
&F(N)=\sum_{e,m\in Z}\left({m^2 N\over{2}}-{e^2\over{2N}}\right)
e^{\pi i\tau(e/\sqrt{N} +m\sqrt{N})^2/2}
e^{-\pi i\btau(e/\sqrt{N} -m\sqrt{N})^2/2}.\cr}\eqn\partii$$
{}From the asymptotic expression of $F(N)$
$$F(N)\sim -{1\over{2\pi\tau_2^{2/3}}}{\sqrt{N}\over{2}}\quad{\rm for}
\quad N>>1,\eqn\eq$$
one can find $\delta_A Z$ and $\delta_D Z$ to be
$$\delta_A Z\cong
-{(q-p)\over{2\sqrt{\tau_2}}}{1\over{|\eta(\tau)|^2}},\qquad
\delta_D Z\cong
-{(q-p)\over{4\sqrt{\tau_2}}}{1\over{|\eta(\tau)|^2}}.\eqn\partdiff$$

Next we compute the one-point function on the torus using the method
used in [\GaS,\Rava].
Since $\Delta[\Phi_{1,3}]\cong 1$, one can express the
one-point function with the characters
$$\eqalign{\big\langle\Phi_{1,3}\big\rangle_{\rm torus}(\tau,\btau)&=
4\pi |z|^2\sum_{F\in {\cal A}}\big\langle\Phi_{F}
\bigm|\Phi_{1,3}(z,\bz)\bigm|\Phi_{F}\big\rangle
\ch_{h}(\tau){\overline\ch}_{\overline h}(\btau),\cr
&=4\pi |z|^2\sum_{F\in {\cal A}}C_{(1,3),F,F}
\ch_{h}(\tau){\overline\ch}_{\overline h}(\btau),\cr}\eqn\onept$$
where ${\cal A}$ denotes a complete set of
primary fields $F$ with conformal dimension $(h,{\overline h})$.
The characters are expressed as a sum of infinite terms as follows:
$$\ch_{r,s}(\tau)={K_{r,s}(\tau)-K_{r,-s}(\tau)\over{\eta(\tau)}},
\quad K_{r,s}(\tau)=\sum_{n=-\infty}^{\infty}\exp\left[2\pi i\tau
{(2pqn+qr-ps)^2\over{4pq}}\right].\eqn\eq$$

Since $p,q>>1$ and $\tau_2>0$, most terms in $K_{r,s}$ vanish
except those with $n=0$ and $qr-ps\sim p,q$.
By the same reason, $K_{r,-s}$ becomes negligible.
Therefore, the character can be approximated as
$$\ch_{r,s}(\tau)\cong {1\over{\eta(\tau)}}\exp\left[\pi i\tau
{(qr-ps)^2\over{2pq}}\right].\eqn\approx$$
The integer $\lambda=qr-ps$ covers all integers between $-pq$ and
$pq$ only once for $1\le r\le p$ and $1\le s\le q$ for $(A,A)$ and all
the odd integers for $(A,D)$.
Furthermore, as pointed out in [\Rava],
the non-diagonal combinations like $\ch_{r,s}{\overline\ch}_{r,q-s}$
are negligible because $|qr-p(q-s)|\sim pq$ if $|qr-ps|\sim {\cal
O}(1)$.

Using these observations and the coefficients of the diagonal three-point
functions
in Eq.\coeff, one can obtain the one-point function on the torus to be
$$\eqalign{\big\langle\Phi_{1,3}\big\rangle_{\rm torus}&\cong
-\alpha{\pi^2\over{\sqrt{3}|\eta|^2}}\sum_{\lambda=-q}^{q}
{\Gamma\left({p+\lambda\over{q}}\right)
\Gamma\left({p-\lambda\over{q}}\right)\over{
\Gamma\left({q-p+\lambda\over{q}}\right)
\Gamma\left({q-p-\lambda\over{q}}\right)}}
\exp\left[-\pi\tau_2{\lambda^2\over{pq}}\right]+{\cal O}(\epsilon)\cr
&\cong\alpha{q\pi^2\over{\sqrt{3}|\eta|^2}}\int_{-\infty}^{\infty}
dx x^2e^{-\pi\tau_2x^2}={\alpha\over{|\eta|^2}}
{q\pi\over{2\sqrt{3}\tau_2^{3/2}}},\cr}\eqn\onepti$$
where $\alpha=1\ {\rm or}\ 1/2$ for the $(A,A)$ and $(A,D)$ MIPFs,
respectively.
Comparing this with Eq.\fixed, one can confirm Eq.\partdiff\ with $X=Y=A$
and $X=Y=D$:
$${\cal M}_{(p,q)}(A,A)\longrightarrow{\cal M}_{(2p-q,p)}(A,A),\qquad
{\cal M}_{(p,q)}(A,D)\longrightarrow{\cal M}_{(2p-q,p)}(A,D).\eqn\final$$

{\bf 4.}
We showed so far that there can exist a RG flow from the $(p,q)$
minimal CFT to the $(2p-q,p)$ minimal CFT due to the least relevent
operator $\Phi_{1,3}$.
Although our argument is rigorous only for $q>>(q-p)$, it may be possible to
extend our results to all possible pairs of $(p,q)$.
For the unitary CFTs, in particular, Al.B. Zamolodchikov showed the RG
flows for any $p$ using the conjectured thermodynamic Bethe
Ansatz (TBA) equations.
\refmark\zami\
For example, the RG flow connects the tricritical Ising model with the
Ising model.
Although the direct derivation of the TBA equations is still missing,
the TBA analysis can support the conjecture that the RG flow exists for
all $(p,p+1)$ unitary CFTs.
Interesting point is that the conjectured TBA equations for the massless
$g>0$ field theories are given by those for the massive
$g<0$ theories, i.e. the restricted sine-Gordon theories with
the massless left- and right-moving particles instead of the massive
particles.
It would be interesting to study the RG flows of the $(p,q)$ minimal
CFTs in the same way as the unitary cases. The conjectured TBA equations
will be provided by those for the massive
theories.\refmark{\AaN,\ahn}\
This TBA analysis of the RG flow of the minimal CFTs based on the TBA
equations will be reported elsewhere.\refmark\AaNi\

One interesting application of the RG flow
is to identify new integrable model which flows into
the `Yang-Lee edge singularity model',
identified with ${\cal M}_{(2,5)}$.\refmark\Cardy\
This new model which is ${\cal M}_{(5,8)}$
can be realized as a UV (short distance) limit of the Yang-Lee model
perturbed by the dimension $4$ operator $T_4(z)$
which is a decendent field of the
vacuumat level $4$:
$$S_{(5,8)}=S_{\rm YL}+g\int d^2 z T_4(z){\overline T}_4(\bz).\eqn\eq$$

Finally, one can think of the RG flows of the $SU(2)$ coset CFTs
which have extended symmetries like the superconformal invariance.
For the unitary CFTs, it has been claimed that there exist the
RG flows into new IR fixed points by the perturbation of the
least relevent operator.\refmark\Coset\
Based on this observation, we can conjecture the
following RG flows of the $SU(2)$ coset non-unitary CFTs due to the
least relevent operator:
$${SU(2)_K\times SU(2)_L\over{SU(2)_{K+L} }}+g\Phi_{\rm pert}
\longrightarrow
{SU(2)_K\times SU(2)_{L-K}\over{SU(2)_L }}+g'\Phi'_{\rm pert},\eqn\eq$$
where $L+2=p/(q-p)$ and $\Delta(\Phi_{\rm pert})=(K+L)/(K+L+2)$ and
$\Delta(\Phi'_{\rm pert})=(K+L+2)/(K+L)$.
Also, interesting is to compare these
massless field theories $(g>0)$ with the massive theories $(g<0)$ considered
in [\ahni] where exact $S$-matrices and particle spectrum are proposed.

{\bf Note} After finishing this paper, we noticed
a preprint [\lassig] which gets a similar result for the diagonal
theories.

\ack

It is a pleasure to thank our collegues at Cornell and S. Nam at Seoul
for various helpful discussions.

%\endpage
{\singlespace\refout}
\end